\documentclass[aps,prc,twocolumn,showpacs,floatfix,nofootinbib]{revtex4}

\usepackage{amsmath,graphicx}
\usepackage[colorlinks=true,linktocpage=true,linkcolor=blue,citecolor=blue]{hyperref}
\usepackage[usenames,dvipsnames]{color}

\begin{document}

\preprint{}
\title{Transport coefficients for bulk viscous evolution in the relaxation-time approximation}
\author{Amaresh Jaiswal}
\affiliation {Tata Institute of Fundamental Research,
Homi Bhabha Road, Mumbai 400005, India}
\author{Radoslaw Ryblewski}
\affiliation {Department of Physics, Kent State University, Kent, Ohio 44242, USA and\\
The H. Niewodnicza\'nski Institute of Nuclear Physics, Polish Academy of Sciences, PL-31342 Krak\'ow, Poland}
\author{Michael Strickland}
\affiliation {Department of Physics, Kent State University, Kent, Ohio 44242, USA}
\date{\today}

\begin{abstract}

We derive the form of the viscous corrections to the phase-space 
distribution function due to the bulk viscous pressure and shear stress tensor 
using the iterative Chapman-Enskog method. We then calculate the 
transport coefficients necessary for the second-order hydrodynamic 
evolution of the bulk viscous pressure and the shear stress tensor. 
We demonstrate that the transport coefficients obtained using the 
Chapman-Enskog method are different than those obtained previously 
using the 14-moment approximation for a finite particle mass. Specializing 
to the case of boost-invariant and transversally homogeneous 
longitudinal expansion, we show that the transport coefficients 
obtained using the Chapman-Enskog method result in better agreement 
with the exact solution of the Boltzmann equation in the relaxation-time 
approximation compared to results obtained in the 14-moment 
approximation. Finally, we explicitly confirm that the time 
evolution of the bulk viscous pressure is significantly affected by 
its coupling to the shear stress tensor.

\end{abstract}



\maketitle

\section{Introduction}

Relativistic hydrodynamics has been quite successful in explaining a 
wide range of collective phenomena observed in astrophysics, 
cosmology, and the physics of high-energy heavy-ion collisions. The 
theory of relativistic hydrodynamics is formulated as a gradient 
expansion where ideal hydrodynamics is the zeroth order. The 
first-order relativistic Navier-Stokes (NS) theory \cite 
{Eckart:1940zz, Landau} leads to acausal signal propagation which is 
rectified in the second-order Israel-Stewart (IS) theory \cite 
{Israel:1979wp}. The derivation of IS equations proceeds in a 
variety of ways \cite {Romatschke:2009im}. For instance, in the 
derivations based on the second law of thermodynamics ($\partial_\mu 
S^\mu \geq 0$), where $S^\mu$ is the generalized entropy 
four-current, the transport coefficients related to relaxation times 
for shear and bulk viscous pressures remain undetermined, and have 
to be obtained from kinetic theory \cite
{Israel:1979wp,Muronga:2003ta}. On the other hand, the derivations 
based on kinetic theory require the non-equilibrium phase-space 
distribution function, $f(x,p)$, to be specified. Consistent and 
accurate determination of the form of the dissipative equations and 
the associated transport coefficients is currently an active 
research area \cite{York:2008rr,El:2009vj,Denicol:2012cn, 
Denicol:2010xn,Jaiswal:2012qm,Jaiswal:2013fc,Bhalerao:2013aha, 
Denicol:2014vaa,Jaiswal:2013npa,Jaiswal:2013vta,Bhalerao:2013pza, 
Romatschke:2003ms,Martinez:2010sc,Martinez:2012tu,Florkowski:2013lza, 
Bazow:2013ifa,Nopoush:2014pfa,Florkowski:2014bba,Florkowski:2014sfa}.

The existence of thermodynamic gradients in a non-equilibrium system 
gives rise to thermodynamic forces which in turn results in various 
transport phenomena. In order to calculate the associated transport 
coefficients, it is convenient to first specify the non-equilibrium 
single particle phase-space distribution function $f(x,p)$. The two most 
commonly used methods to determine the form of $f(x,p)$ when the 
system is close to local thermodynamic equilibrium are (1) Grad's 
14-moment approximation \cite{Grad} and (2) the Chapman-Enskog 
method \cite{Chapman}. While Grad's moment method has been widely 
used in the formulation of causal relativistic dissipative 
hydrodynamics from kinetic theory \cite{Israel:1979wp, 
Muronga:2003ta,El:2009vj,Denicol:2010xn,Denicol:2012cn, 
Jaiswal:2012qm,Jaiswal:2013fc,Bhalerao:2013aha,Denicol:2014vaa, 
Romatschke:2009im}, the Chapman-Enskog method remains less explored 
\cite{York:2008rr,Jaiswal:2013npa, Jaiswal:2013vta}. Although both 
the methods involve expanding $f(x,p)$ around the equilibrium 
distribution function $f_0(x,p)$, in Refs.~\cite 
{Jaiswal:2013npa,Jaiswal:2013vta} it was demonstrated that the 
Chapman-Enskog method in the relaxation-time approximation (RTA) 
gives better agreement with both microscopic Boltzmann simulations 
and exact solutions of the RTA Boltzmann equation. This seems to 
stem from the fact that the Chapman-Enskog method does not require a 
fixed-order Grad's-moment expansion.

Relativistic viscous hydrodynamics has been used extensively to 
study and understand the evolution of the strongly interacting, hot 
and dense matter created in high-energy heavy-ion collisions; see 
Ref. \cite{Heinz:2013th} for a recent review. While much of the 
research on this topic is devoted to the extraction of the shear 
viscosity to entropy density ratio $\eta/s$ from the analysis of the 
flow data \cite{Romatschke:2007mq,Song:2010mg,Schenke:2011bn}, a 
systematic and self-consistent study of the effect of bulk viscosity 
in numerical simulations of heavy-ion collisions has not been 
performed. The relative lack of effort in this direction may be 
attributed to the fact that the bulk viscosity of hot QCD matter is 
estimated to be much smaller compared to the shear viscosity. 
However, it is important to note that for the range of temperature 
probed experimentally in heavy-ion collisions, the magnitude and 
temperature dependence of bulk viscosity is unknown \cite 
{Moore:2008ws,Noronha-Hostler:2014dqa} and could be large enough to 
affect the spatio-temporal evolution of the QCD matter. Moreover, 
since QCD is a non-conformal field theory, bulk viscous corrections 
to the energy momentum tensor should not be neglected in order to 
correctly understand the dynamics of a QCD system.  

From a theoretical perspective, the second-order transport 
coefficients that appear in the evolution equation for the bulk viscous 
pressure are less understood compared to those of the shear stress 
tensor. In Refs. \cite{Jaiswal:2013fc,Bhalerao:2013aha}, it was 
shown that the relaxation time for bulk viscous evolution can be 
obtained by employing the second law of thermodynamics in a kinetic 
theory set up. While for finite masses, the transport coefficients 
corresponding to bulk viscous pressure and shear stress tensor have 
been explicitly obtained by employing the 14-moment approximation 
\cite{Denicol:2014vaa,Denicol:2014mca}, they still remain to be 
determined using the Chapman-Enskog method. In this paper, we calculate 
the transport coefficients appearing in the second-order viscous 
evolution equations for non-vanishing masses using the method of 
Chapman-Enskog expansion. We compare the mass dependence of these 
coefficients with those obtained using the 14-moment approximation. In 
the case of one-dimensional scaling expansion of the viscous medium, 
we demonstrate that our results are in better agreement with the 
exact solution of the massive (0+1)-dimensional Boltzmann equation in the relaxation 
time approximation \cite{Florkowski:2014sfa} than to those obtained using 
the 14-moment approximation. We also confirm that generation of bulk 
viscous pressure is affected more by its coupling to the shear stress 
tensor than the first-order expansion rate of the system, in agreement
with Ref.~\cite{Denicol:2014mca}.


\section{Relativistic hydrodynamics}

The hydrodynamic evolution of a system having no net conserved 
charges (vanishing chemical potential) is governed by the local 
conservation of energy and momentum, $\partial_\mu T^{\mu\nu}=0$. 
The energy-momentum tensor, $T^{\mu\nu}$, characterizing the 
macroscopic state of a system, can be expressed in terms of a 
single-particle phase-space distribution function and tensor 
decomposed into hydrodynamic degrees of freedom \cite{deGroot},
\begin{equation}\label{NTD}
T^{\mu\nu} = \!\int\! dP \, p^\mu p^\nu f(x,p) = \epsilon u^\mu u^\nu 
- (P+\Pi)\Delta^{\mu\nu} + \pi^{\mu\nu}.
\end{equation}
Here $dP\equiv g d^3p/[(2 \pi)^3p^0]$ is the invariant momentum-space 
integration measure, where $g$ is the 
degeneracy factor, $p^\mu$ is the particle four-momentum, and 
$f(x,p)$ is the phase-space distribution function. In the tensor 
decomposition, $\epsilon$, $P$, $\Pi$, and $\pi^{\mu\nu}$ are energy 
density, thermodynamic pressure, bulk viscous pressure, and shear 
stress tensor, respectively. The projection operator 
$\Delta^{\mu\nu}\equiv g^{\mu\nu}-u^\mu u^\nu$ is orthogonal to the 
hydrodynamic four-velocity $u^\mu$ defined in the Landau frame: 
$T^{\mu\nu} u_\nu=\epsilon u^\mu$. The metric tensor is Minkowskian, 
$g^{\mu\nu}\equiv\mathrm{diag}({+}1,{-}1,{-}1,{-}1)$.

The projection of $\partial_\mu T^{\mu\nu}=0$ along and orthogonal 
to $u^\mu$ leads to the evolution equations for $\epsilon$ and 
$u^\mu$,
\begin{align}\label{evol}
\dot\epsilon + (\epsilon+P+\Pi)\theta - \pi^{\mu\nu}\sigma_{\mu\nu} &= 0, \\
(\epsilon+P+\Pi)\dot u^\alpha - \nabla^\alpha (P+\Pi) + \Delta^\alpha_\nu \partial_\mu \pi^{\mu\nu}  &= 0.
\end{align}
Here we have used the standard notation $\dot A\equiv 
u^\mu\partial_\mu A$ for the co-moving derivative, 
$\theta\equiv\partial_\mu u^\mu$ for the expansion scalar, 
$\sigma^{\mu\nu}\equiv \frac{1}{2}(\nabla^\mu u^\nu + \nabla^\nu u^\mu) 
-\frac{1}{3} \theta \Delta^{\mu\nu}$ for the velocity stress tensor, and 
$\nabla^\alpha\equiv\Delta^{\mu\alpha}\partial_\mu$ for space-like 
derivatives. The inverse temperature, $\beta\equiv1/T$, is 
determined by the matching condition $\epsilon=\epsilon_0$ where 
$\epsilon_0$ is the equilibrium energy density. In terms of the 
equilibrium distribution function $f_0$, the energy density and the 
thermodynamic pressure can be written as
\begin{eqnarray}\label{EDTP1}
\epsilon_0 &=& u_\mu u_\nu \!\int\! dP \, p^\mu p^\nu f_0, 
\\ \label{EDTP2}
P_0 &=& -\frac{1}{3}\Delta_{\mu\nu} \!\int\! dP \, p^\mu p^\nu f_0,
\end{eqnarray}
respectively.
For a classical Boltzmann gas with vanishing chemical 
potential, the equilibrium distribution function is given by 
$f_0=\exp(-\beta\,u\cdot p)$ where $u \cdot p \equiv u_\mu p^\mu$.

From Eqs.~(\ref{EDTP1}) and (\ref{EDTP2}) one obtains $\dot\epsilon$ and $\nabla^\alpha P$
in terms of derivatives of $\beta$ as
\begin{equation}\label{DEdP}
\dot\epsilon = -I_{30}^{(0)}\dot\beta, \quad
\nabla^\alpha P = I_{31}^{(0)}\nabla^\alpha\beta,
\end{equation}
where  
\begin{equation}\label{IC}
I_{nq}^{(r)} \equiv \frac{1}{(2q+1)!!}\!\int\! dP\, 
(u\cdot p)^{n-2q-r}\,(\Delta_{\mu\nu}p^\mu p^\nu)^q f_0.
\end{equation}
Here we readily identify $I_{20}^{(0)}=\epsilon$ and 
$I_{21}^{(0)}=-P$. The integrals $I_{nq}^{(r)}$ satisfy the 
following relations
\begin{align}
I_{nq}^{(r)}& = I_{n-1,q}^{(r-1)} ~~{\rm for}~ n>2q, \label{prop1} \\
I_{nq}^{(r)} & = \frac{1}{(2q+1)}\left[m^2 I_{n-2,q-1}^{(r)} - I_{n,q-1}^{(r)}\right], \label{prop2} \\
I_{nq}^{(0)}&= \frac{1}{\beta}\left[-I_{n-1,q-1}^{(0)} + (n-2q)I_{n-1,q}^{(0)} \right]. \label{prop3}
\end{align}
The above relations lead to the following identities
\begin{eqnarray}\label{identities1}
I_{31}^{(0)}&=&-\frac{1}{\beta}(\epsilon+P), 
\\\label{identities11}
I_{30}^{(0)}&=&\frac{1}{\beta}[3\epsilon+(3+z^2)P],
\end{eqnarray}
where $z\equiv\beta m$ with $m$ being the mass of the particle. 
Substituting the expressions for $\dot\epsilon$ and $\nabla^\alpha P$
from Eq. (\ref{DEdP}) in Eq. (\ref{evol}), one obtains
\begin{align}
\dot\beta &= \frac{\beta(\epsilon+P)}{3\epsilon+(3+z^2)P}\theta 
+ \frac{\beta(\Pi\theta - \pi^{\rho\gamma}\sigma_{\rho\gamma})}{3\epsilon+(3+z^2)P}, \label{evol1} \\
\nabla^\alpha\beta &= -\beta\dot u^\alpha - \frac{\beta}{\epsilon+P} 
\left(\Pi\dot u^\alpha - \nabla^\alpha\Pi + \Delta^\alpha_\nu\partial_\mu\pi^{\mu\nu}\right). \label{evol2}
\end{align}
The above identities are used later to obtain the form of viscous 
corrections to the distribution function and derive evolution 
equations for shear and bulk viscous pressures.

Close to local thermodynamic equilibrium, the phase-space 
distribution function can be written as $f=f_0+\delta f$, where 
$\delta f\ll f$. From Eq. (\ref {NTD}), the bulk viscous pressure 
$\Pi$ and the shear stress tensor $\pi^{\mu\nu}$ can be expressed in 
terms of the non-equilibrium part of the distribution function 
$\delta f$ as \cite{deGroot}
\begin{align}
\Pi &= -\frac{1}{3}\Delta_{\alpha\beta} \!\int\! dP \, p^\alpha p^\beta\, \delta f, \label{BVP}\\
\pi^{\mu\nu} &= \Delta^{\mu\nu}_{\alpha\beta} \!\int\! dP \, p^\alpha p^\beta\, \delta f, \label{SST}
\end{align}
where $\Delta^{\mu\nu}_{\alpha\beta}\equiv \frac{1}{2}
(\Delta^{\mu}_{\alpha}\Delta^{\nu}_{\beta} + 
\Delta^{\mu}_{\beta}\Delta^{\nu}_{\alpha}) - 
\frac{1}{3} \Delta^{\mu\nu}\Delta_{\alpha\beta}$ is a traceless symmetric 
projection operator orthogonal to $u^\mu$. In the following, we 
iteratively solve the RTA Boltzmann equation to obtain $\delta f$ up to first order.


\section{Viscous evolution equations}

The relativistic Boltzmann equation in the RTA is given by \cite{Anderson_Witting},
\begin{equation}\label{RBE}
p^\mu\partial_\mu f = -\left(u\cdot p\right) \frac{\delta f}{\tau_{\rm eq}},
\end{equation}
where $\tau_{\rm eq}$ is the relaxation time. To ensure the straightforward 
conservation of particle current and energy-momentum tensor, 
$\tau_{\rm eq}$ should be independent of momenta and $u^\mu$ should 
be defined in the Landau frame \cite{Anderson_Witting}. Rewriting 
Eq.~(\ref{RBE}) in the form $f=f_0-(\tau_{\rm eq}/u\cdot 
p)\,p^\mu\partial_\mu f$ and solving iteratively, one obtains \cite
{Romatschke:2011qp,Jaiswal:2013npa}
\begin{eqnarray}\label{F1F21}
f_1 &=& f_0 -\frac{\tau_{\rm eq}}{u\cdot p} \, p^\mu \partial_\mu f_0,
\\ \label{F1F22}
f_2 &=& f_0 -\frac{\tau_{\rm eq}}{u\cdot p} \, p^\mu \partial_\mu f_1,
\\ \nonumber 
&\vdots&
\end{eqnarray}
where $f_n=f_0+\delta f^{(1)}+\delta f^{(2)}+\cdots+\delta f^{(n)}$.
To first-order in derivatives, we have
\begin{equation}\label{FOC1}
\delta f^{(1)} = -\frac{\tau_{\rm eq}}{u\cdot p} \, p^\mu \partial_\mu f_0.
\end{equation}
Using Eqs. (\ref{evol1}) and (\ref{evol2}) and consistently ignoring 
higher order gradient correction terms, one obtains \cite{Romatschke:2011qp}
\begin{equation}\label{FOC}
\delta f = \frac{\beta\tau_{\rm eq}}{u\cdot p}\left[ \frac{1}{3}\left\{m^2-(1-3c_s^2)(u\cdot p)^2\right\}\theta 
+ p^\mu p^\nu\sigma_{\mu\nu} \right]f_0.
\end{equation}
Here, the velocity of sound squared $c_s^2\equiv dP/d\epsilon$ can 
be expressed as
\begin{equation}\label{cs2}
c_s^2 = \frac{\epsilon+P}{3\epsilon+(3+z^2)P}.
\end{equation}
We observe that the above expression reduces to $c_s^2=1/3$ in the 
ultra-relativistic ($z\to0$) limit.

Substituting Eq. (\ref{FOC1}) in Eqs. (\ref{BVP}) and (\ref{SST}), 
one obtains
\begin{eqnarray}\label{FOE1}
\Pi &=& -\tau_{\rm eq}\beta_\Pi\theta , 
\\\label{FOE2}
\pi^{\mu\nu} &=& 2\tau_{\rm eq}\beta_\pi\sigma^{\mu\nu} ,
\end{eqnarray}
where
\begin{eqnarray}\label{FOTC1}
\beta_\Pi &=& \frac{5}{3}\beta\, I_{42}^{(1)} - (\epsilon+P)c_s^2, 
\\ \label{FOTC2}
\beta_\pi &=& \beta\, I_{42}^{(1)}.
\end{eqnarray}
Replacing the velocity gradients appearing in Eq. (\ref{FOC}) with viscous 
pressures using Eqs. (\ref{FOE1}) and (\ref{FOE2}), one obtains
\begin{align}\label{deltaf}
\delta f =\,& - \frac{\beta f_0}{3(u\cdot p)\beta_\Pi}\left[m^2-(1-3c_s^2)(u\cdot p)^2\right]\Pi \nonumber\\
&\,+ \frac{\beta f_0}{2(u\cdot p)\beta_\pi}\;p^\mu p^\nu\pi_{\mu\nu}.
\end{align}
The above form of $\delta f$ is analogous to the 14-moment 
approximation and can be used in Cooper-Frye prescription for 
particle production \cite{Cooper:1974mv}.

To obtain second-order evolution equations for the bulk viscous pressure 
and the shear stress tensor, we follow the methodology discussed in Ref. 
\cite{Denicol:2010xn}. We express the evolution of bulk viscous 
pressure and shear stress tensor given in Eqs. (\ref {BVP}) and (\ref
{SST}) as
\begin{align}
\dot\Pi &= -\frac{1}{3}\Delta_{\alpha\beta} \!\int\! dP \, p^\alpha p^\beta \delta\dot f, \label{SBE}\\
\dot\pi^{\langle\mu\nu\rangle} &= \Delta^{\mu\nu}_{\alpha\beta} \!\int\! dP \, p^\alpha p^\beta \delta\dot f , \label{SSE}
\end{align}
respectively.
The comoving derivative $\delta\dot f$ can be obtained by 
rewriting Eq. (\ref{RBE}) in the form
\begin{equation}\label{DFD}
\delta\dot f = -\dot f_0 - \frac{1}{u\cdot p}p^\gamma\nabla_\gamma f - \frac{\delta f}{\tau_{\rm eq}}.
\end{equation}
Using the above expression for $\delta\dot f$ in Eqs. (\ref{SBE}) 
and (\ref{SSE}), one obtains
\begin{align} 
\dot\Pi  =& - \frac{\Pi}{\tau_{\rm eq}}
+ \frac{\Delta_{\alpha\beta}}{3}\!\int\! dP \, p^\alpha p^\beta\!\left(\dot f_0 + \frac{1}{u\cdot p}\,
p^\gamma\nabla_\gamma f\right), \label{SOBE}\\
\dot\pi^{\langle\mu\nu\rangle} =& - \frac{\pi^{\mu\nu}}{\tau_{\rm eq}}
- \Delta^{\mu\nu}_{\alpha\beta}\!\int\! dP \, p^\alpha p^\beta\!\left(\dot f_0 + \frac{1}{u\cdot p}\,
p^\gamma\nabla_\gamma f\right). \label{SOSE}
\end{align}
It is clear from Eqs.~(\ref{SOBE}) and (\ref{SOSE}) that there is 
only one time scale to describe the relaxation of the viscous 
evolution equations, i.e., $\tau_{\rm eq}=\tau_\Pi=\tau_\pi$. This 
stems from the fact that the RTA collision term in the Boltzmann 
equation (\ref{RBE}) does not entirely 
capture the microscopic interactions. However, comparing the 
first-order equation, Eqs. (\ref {FOE1}) and (\ref {FOE2}) with the relativistic 
Navier-Stokes equations for bulk and shear pressures, 
$\Pi=-\zeta\theta$ and $\pi^{\mu\nu}=2\eta\sigma^{\mu\nu}$, we 
obtain $\tau_\Pi=\zeta/\beta_\Pi$ and $\tau_\pi=\eta/\beta_\pi$. The 
first-order transport coefficients $\zeta$ and $\eta$ can be 
calculated independently, by taking into account the full microscopic 
behavior of the system. 

Substituting $\delta f$ from Eq. (\ref{deltaf}) in Eqs. (\ref{SOBE}) 
and (\ref{SOSE}) and performing the integrations, one obtains the 
second-order evolution equations for the bulk viscous pressure and shear 
stress tensor
\begin{align}
\dot{\Pi} =& -\frac{\Pi}{\tau_{\Pi}}
-\beta_{\Pi}\theta 
-\delta_{\Pi\Pi}\Pi\theta
+\lambda_{\Pi\pi}\pi^{\mu\nu}\sigma_{\mu \nu }, \label{BULK}\\
\dot{\pi}^{\langle\mu\nu\rangle} =& -\frac{\pi^{\mu\nu}}{\tau_{\pi}}
+2\beta_{\pi}\sigma^{\mu\nu}
+2\pi_{\gamma}^{\langle\mu}\omega^{\nu\rangle\gamma}
-\tau_{\pi\pi}\pi_{\gamma}^{\langle\mu}\sigma^{\nu\rangle\gamma}  \nonumber \\
&-\delta_{\pi\pi}\pi^{\mu\nu}\theta 
+\lambda_{\pi\Pi}\Pi\sigma^{\mu\nu}, \label{SHEAR}
\end{align}
where $\omega ^{\mu \nu }\equiv \frac{1}{2}(\nabla^{\mu}u^{\nu }-\nabla^{\nu }u^{\mu })$ 
is the vorticity tensor.  The transport coefficients appearing above are
\begin{align}\label{coeff1}
\delta_{\Pi\Pi} &= -\frac{5}{9}\,\chi - c_s^2,
\\\label{coeff2}
\lambda_{\Pi\pi} &= 
\frac{\beta}{3\beta_\pi}\!\left(7I_{63}^{(3)}+2I_{42}^{(1)}\right) - c_s^2, 
\\\label{coeff3}
\tau_{\pi\pi} &= 2 + \frac{4\beta}{\beta_\pi}\,I_{63}^{(3)}, 
\\\label{coeff4}
\delta_{\pi\pi} &= \frac{5}{3} + \frac{7\beta}{3\beta_\pi}\,I_{63}^{(3)}, 
\\\label{coeff5}
\lambda_{\pi\Pi} &= -\frac{2}{3}\chi, 
\end{align}
where
\begin{equation}
\chi = \frac{\beta}{\beta_\Pi}\!\left[(1-3c_s^2)\!\left(I_{42}^{(1)}+I_{31}^{(0)}\right) 
- m^2\!\left(I_{42}^{(3)}+I_{31}^{(2)}\right)\right]. 
\end{equation}
Apart from $I_{31}^{(0)}=-(\epsilon+P)/\beta$, see Eq. (\ref 
{identities1}), we need to determine the integrals 
$I_{63}^{(3)}$, $I_{42}^{(1)}$, $I_{42}^{(3)}$, and $I_{31}^{(2)}$. 
In the following, we obtain expressions for these quantities in 
terms of modified Bessel functions of the second kind.


\section{Transport coefficients}

The transport coefficients obtained in the previous section can be 
expressed in terms of modified Bessel functions of the second kind. 
We start from the integral representation of the corresponding 
Bessel function, 
\begin{equation}\label{Bessel}
K_n(z) = \!\int_0^\infty\! d\theta \cosh(n\theta)\, \exp(-z\cosh\theta).
\end{equation}
Using the above form of the Bessel 
function, one obtains the following identities
\begin{align}\label{identities2}
\int_0^\infty\! d\theta \cosh^5\theta\, \exp(-z\cosh\theta) &= 
\frac{1}{16}\left[K_5+5K_3+10K_1\right], \\
\int_0^\infty\! d\theta \cosh^3\theta\, \exp(-z\cosh\theta) &= 
\frac{1}{4}\left[K_3+3K_1\right],\label{identities22}
\end{align}
where the $z$-dependence of $K_n$ is implicitly understood.

\begin{figure}[t]
\begin{center}
\includegraphics[width=\linewidth]{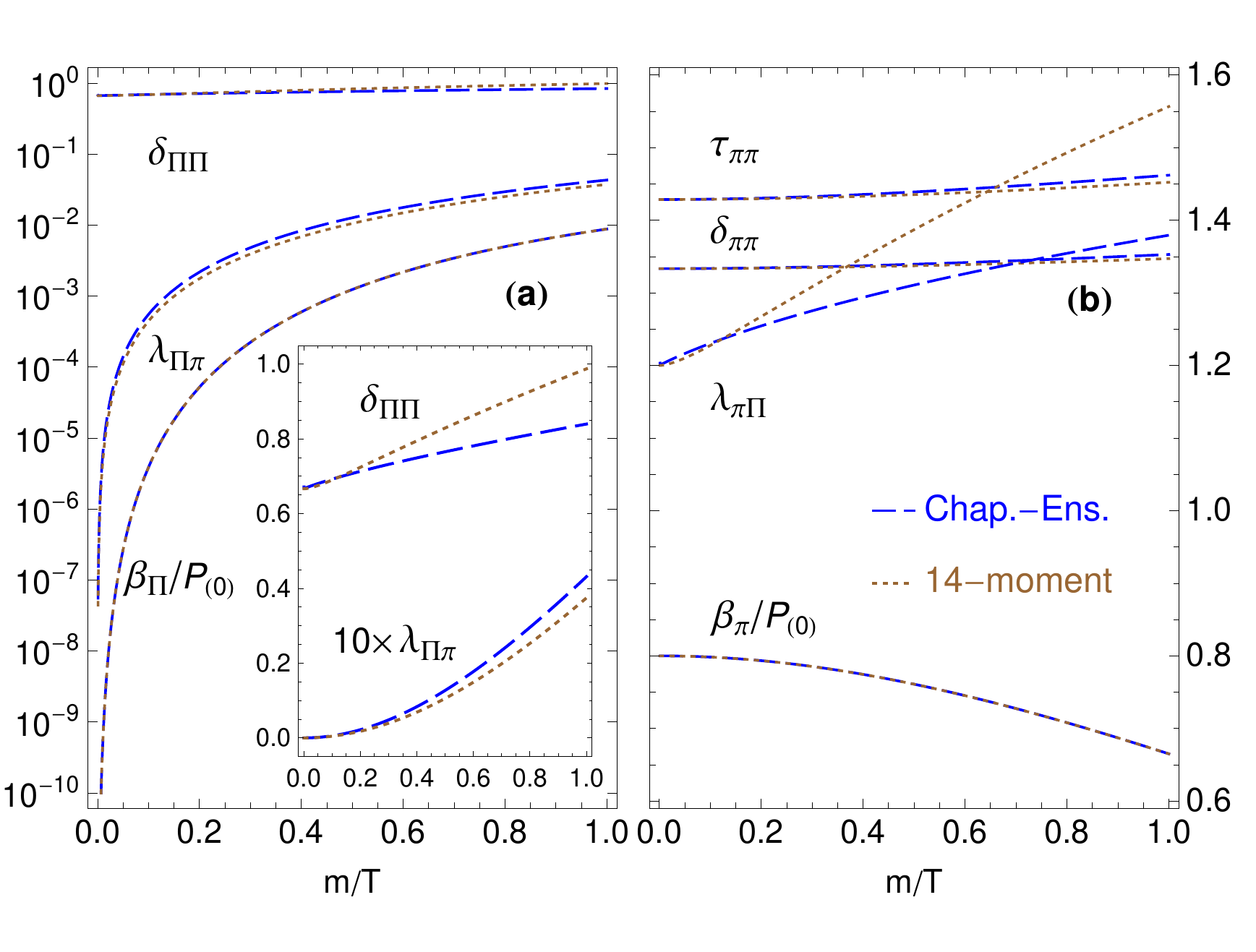}
\end{center}
\vspace{-0.8cm}
\caption{(Color online) Comparison of the exact transport 
	coefficients obtained herein using the Chapman-Enskog method 
	(blue dashed line) with those calculated using the 14-moment 
	approximation (brown dotted line). The two panels correspond to 
	the transport coefficients which enter (a) the bulk viscous 
	pressure and (b) the shear stress tensor evolution equations, as 
	a function of the ratio of mass and temperature. The inset in 
	panel (a) shows the $m/T$-dependence of the transport 
	coefficients $\delta_{\Pi\Pi}$ and $\lambda_{\Pi\pi}$ obtained 
	using the two methods on a linear scale. Here $P_{(0)}$ is the 
	pressure at vanishing mass, i.e., $P_{(0)} \equiv P(m=0,T)$.}
\label{fig_coef}
\end{figure}

The thermodynamic integrals $I_{nq}^{(r)}$ can be cast in a similar 
form,
\begin{align}\label{Inqr}
I_{nq}^{(r)} =&\ \frac{g\,T^{n+2-r}z^{n+2-r}}{2\pi^2(2q+1)!!}(-1)^q\!\int_0^\infty\! d\theta\, 
(\cosh\theta)^{n-2q-r}\nonumber\\
&\qquad\qquad\quad\times(\sinh\theta)^{2q+2}\,\exp(-z\cosh\theta).
\end{align}
By using the identity $\cosh^2\theta-\sinh^2\theta=1$, the integral 
in $I_{nq}^{(r)}$ can be expressed in terms of 
$\cosh\theta$ only. Employing Eqs. (\ref{identities2}) and (\ref{identities22}), one obtains
\begin{align}\label{relint}
I_{63}^{(3)} &= -\frac{gT^5z^5}{210\pi^2}\!\left[\!\frac{1}{16}(K_5\!-11K_3+58K_1)-4K_{i,1}+K_{i,3}\!\right]\!, \\
I_{42}^{(1)} &= \frac{gT^5z^5}{30\pi^2}\left[\frac{1}{16}(K_5-7K_3+22K_1)-K_{i,1}\right], \\
I_{42}^{(3)} &= \frac{gT^3z^3}{30\pi^2}\left[\frac{1}{4}(K_3-9K_1)+3K_{i,1}-K_{i,3}\right], \\
I_{31}^{(2)} &= -\frac{gT^3z^3}{6\pi^2}\left[\frac{1}{4}(K_3-5K_1)+K_{i,1}\right].
\end{align}
Here the function $K_{i,n}$ is defined by the integral
\begin{equation}\label{kin}
K_{i,n}(z) = \!\int_0^\infty\! \frac{d\theta}{(\cosh\theta)^n}\,\exp(-z\cosh\theta),
\end{equation}
which has the following property
\begin{equation}\label{kinkn1}
\frac{d}{dz}K_{i,n}(z) = -K_{i,n-1}(z).
\end{equation}
This identity can also be written in integral form as
\begin{equation}\label{kinkn2}
K_{i,n}(z) = K_{i,n}(0) - \!\int_0^z\! K_{i,n-1}(z') dz'.
\end{equation}
We observe that, by using the series expansion of $K_{i,0}(z)=K_0(z)$, 
the above recursion relation can be employed to evaluate $K_{i,n}(z)$
up to any given order in $z$.

\begin{figure}[t]
\begin{center}
~~~~~~\includegraphics[width=\linewidth]{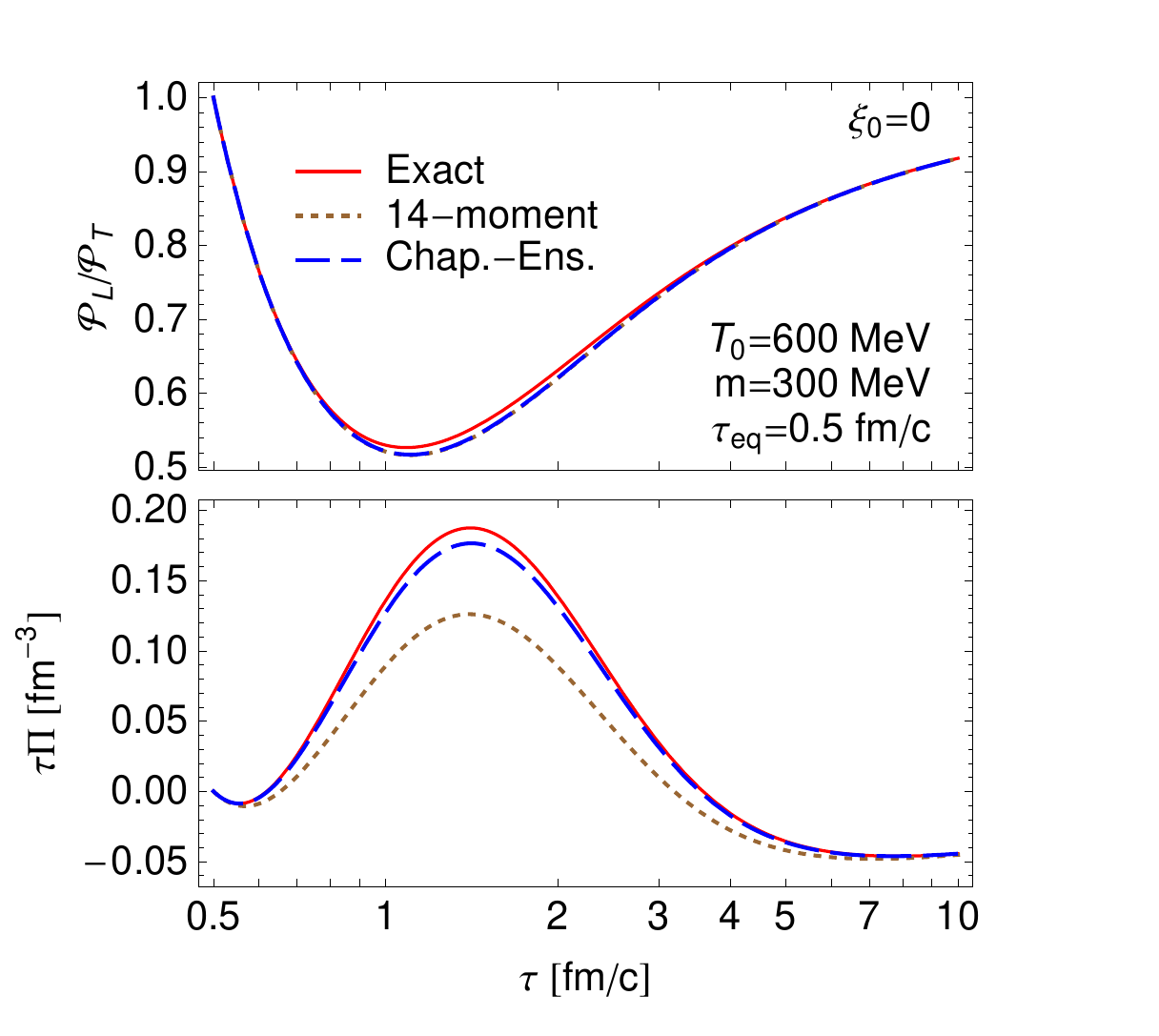}
\end{center}
\vspace{-0.8cm}
\caption{(Color online) Time evolution of the pressure anisotropy 
	${\cal P}_L/{\cal P}_T$ (top) and the bulk viscous pressure 
	times $\tau$ (bottom) for three different calculations: the 
	exact solution of the RTA Boltzmann equation \cite
	{Florkowski:2014sfa} (red solid line), second-order viscous 
	hydrodynamics using the 14-moment method \cite{Denicol:2014vaa} 
	(brown dotted line), and the Chapman-Enskog method used herein 
	(blue dashed line). For both panels we use $T_0=600$ MeV at 
	$\tau_0=0.5$ fm/$c$, $m=300$ MeV, and $\tau_{\rm 
	eq}=\tau_\pi=\tau_\Pi=0.5$ fm/$c$. The initial spheroidal 
	anisotropy in the distribution function, $\xi_0=0$, corresponds 
	to isotropic initial pressures with $\pi_0=0$ and $\Pi_0=0$.} 
\label{fig_0_03}
\end{figure}

\begin{figure}[t]
\begin{center}
\includegraphics[width=\linewidth]{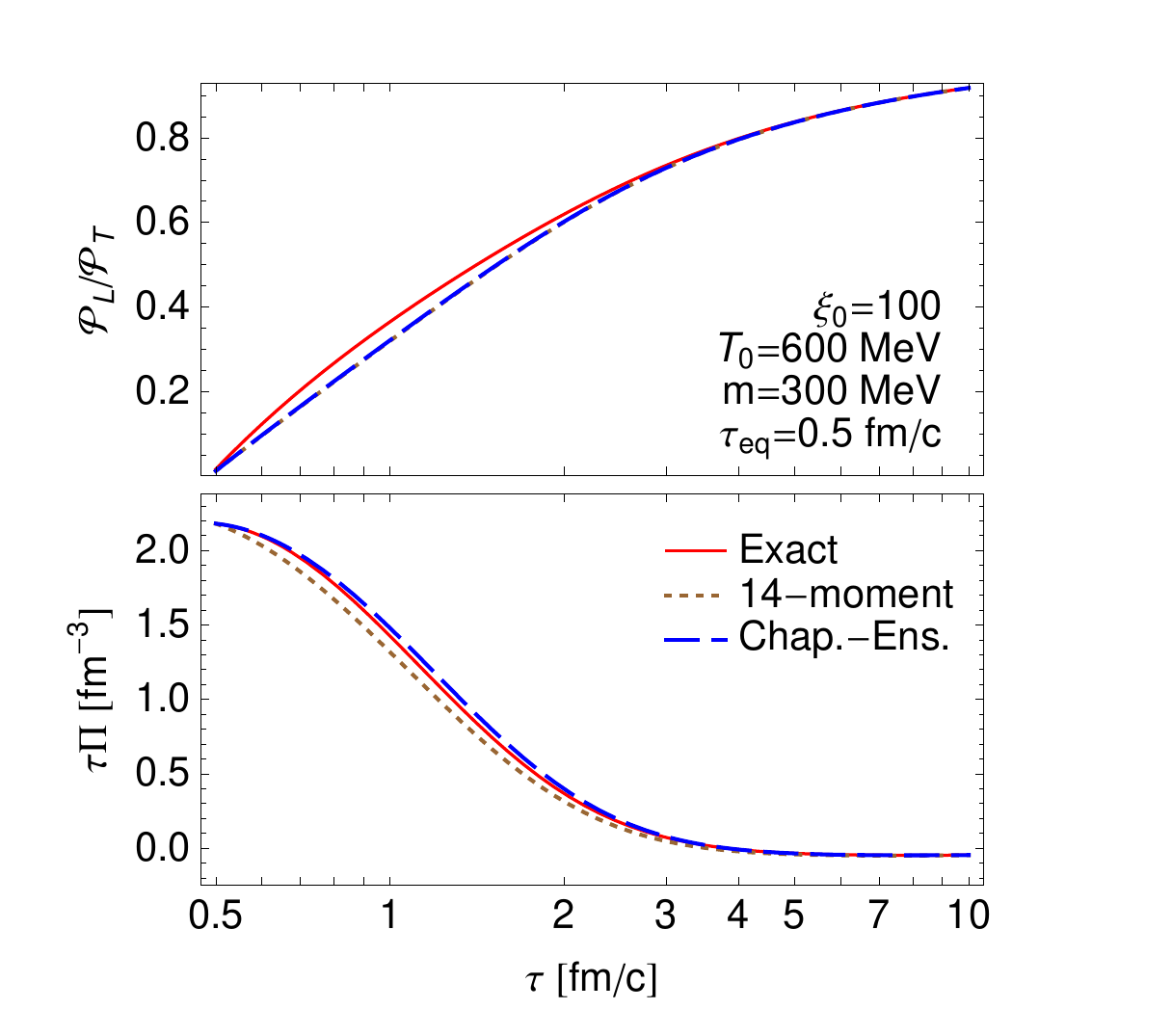}
\end{center}
\vspace{-0.8cm}
\caption{(Color online) Same as Fig. \ref{fig_0_03} except here we 
	take $\xi_0=100$ corresponding to $\pi_0=51.11$ GeV/fm$^3$ and $\Pi_0=0.85$ GeV/fm$^3$.} 
\label{fig_100_03}
\end{figure}
In the results section, we will use the exact expressions for the 
various transport coefficients.  However, before proceeding to the 
numerical results it is possible to compare the analytic small-mass 
expansions of the transport coefficients with the results obtained 
using the 14-moment approximation. With this in mind, we now present 
small-mass expansions of the kinetic coefficients obtained in Eqs. 
(\ref{FOTC1}), (\ref{FOTC2}), and (\ref{coeff1})-(\ref{coeff5}). We 
begin by noting that the quantity $\chi$ that appears in the 
transport coefficients (\ref{coeff1})-(\ref{coeff5}) has the 
following small-mass expansion 

\begin{equation}\label{chiz}
\chi = -\frac{9}{5} - \frac{9\pi z}{50} + {\cal O}(z^2\ln z).
\end{equation}
The small-mass expansions of the transport coefficients entering the bulk evolution equation are
\begin{align}\label{bulkz}
\frac{\beta_\Pi}{\epsilon+P} &= \frac{5z^4}{432} + {\cal O}(z^5), \nonumber\\
\delta_{\Pi\Pi} &= \frac{2}{3} + \frac{\pi z}{10} + {\cal O}(z^2\ln z), \nonumber\\
\lambda_{\Pi\pi} &= \frac{z^2}{18} - \frac{5z^4}{144} + {\cal O}(z^5).
\end{align}
Similarly, the small-mass expansions of the transport coefficients entering the shear tensor evolution equation 
are
\begin{align}\label{shearz}
\frac{\beta_\pi}{\epsilon+P} &= \frac{1}{5} - \frac{z^2}{60} + \frac{z^4}{96} + {\cal O}(z^5), \nonumber\\
\delta_{\pi\pi} &= \frac{4}{3} + \frac{z^2}{36} - \frac{25z^4}{864} + {\cal O}(z^5), \nonumber\\
\tau_{\pi\pi} &= \frac{10}{7} + \frac{z^2}{21} - \frac{25z^4}{504} + {\cal O}(z^5), \nonumber\\
\lambda_{\pi\Pi} &= \frac{6}{5} + \frac{3\pi z}{25} + {\cal O}(z^2\ln z).
\end{align}
We observe that while the expressions for $\beta_\Pi$ and $\beta_\pi$
in Eqs. (\ref{bulkz}) and (\ref{shearz}) are identical to those 
obtained by using the 14-moment method \cite{Denicol:2014vaa, 
Denicol:2014mca}, the other coefficients agree only up to the 
constant term in their respective Taylor expansions in powers of $z$.

Having established that the Chapman-Enskog transport coefficients 
are different than the 14-moment transport coefficients even for 
small masses, we now turn to the exact numerical evaluation of the 
transport coefficients for arbitrary mass. In Fig.~\ref{fig_coef} we 
compare the exact transport coefficients obtained herein using the 
Chapman-Enskog method (blue dashed line) with those calculated using 
the 14-moment approximation (brown dotted line). Figure \ref
{fig_coef} (a) and (b) shows the transport coefficients entering the 
evolution equations for the bulk viscous pressure and the shear 
stress tensor, respectively. In the inset of Fig. \ref{fig_coef} 
(a), we show the $m/T$ dependence of the transport coefficients 
$\delta_{\Pi\Pi}$ and $\lambda_{\Pi\pi}$ (multiplied by a factor of 
10) obtained using the two methods on a linear scale. We observe 
that the two methods lead to very similar values of the transport 
coefficients for small values of $z=m/T$.  For large values of $z$, 
the differences are significant for some transport coefficients. For 
example, at $z=1$, the values of $\lambda_{\pi\Pi}$, 
$\delta_{\Pi\Pi}$ and $\lambda_{\Pi\pi}$ in the two cases differ by 
approximately $15\%$, $20\%$ and $25\%$, respectively.

Another quantity of interest is the square of the sound velocity 
in the medium, $c_s^2$, which for small masses is approximately
\begin{equation}\label{cs2z}
\frac{1}{3} - c_s^2 = \frac{z^2}{36} - \frac{5z^4}{864} + {\cal O}(z^6\ln z).
\end{equation}
In the RTA, by comparing the relativistic NS equations, 
$\Pi=-\zeta\theta$ and $\pi^{\mu\nu}= 2\eta\sigma^{\mu\nu}$, with 
Eqs. (\ref{FOE1}) and (\ref{FOE2}), one obtains $\zeta/\eta= 
\beta_\Pi/\beta_\pi$. Using the series expansion in $z$, one obtains
\begin{equation}\label{zetabyeta}
\frac{\zeta}{\eta} = 75\left(\frac{1}{3}-c_s^2\right)^2 + {\cal O}(z^5).
\end{equation}
The relation in Eq. (\ref{zetabyeta}) can also be obtained by using 
the expressions for $\zeta$ and $\eta$ presented in Ref. \cite
{Florkowski:2014sfa}.\footnote{We note that the factor $75$ is different than 
the value obtained in Ref. \cite{Denicol:2014vaa}, which was $72.75$.}
It is interesting to note that the form of the above expression is 
similar to the well known relation, $\zeta/\eta=15(1/3-c_s^2)^2$, 
derived by Weinberg \cite{Weinberg}. However, we find the proportionality 
constant to be exactly five times larger than that obtained by Weinberg.
    
\begin{figure}[t]
\begin{center}
\includegraphics[width=\linewidth]{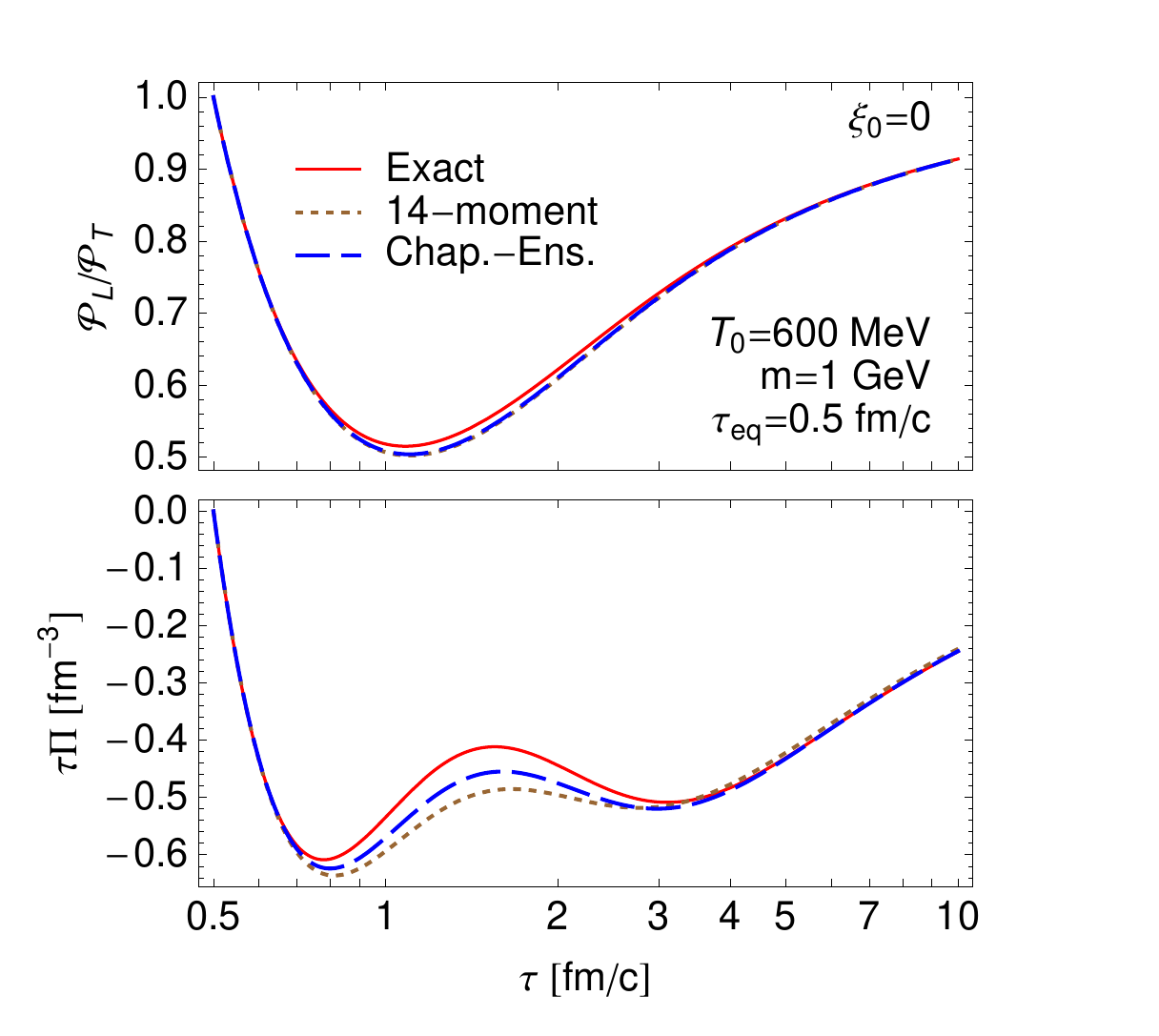}
\end{center}
\vspace{-0.8cm}
\caption{(Color online) Same as Fig. \ref{fig_0_03} except here 
	we take $m=1$ GeV.} 
\label{fig_0_1}
\end{figure}

\begin{figure}[t]
\begin{center}
\includegraphics[width=\linewidth]{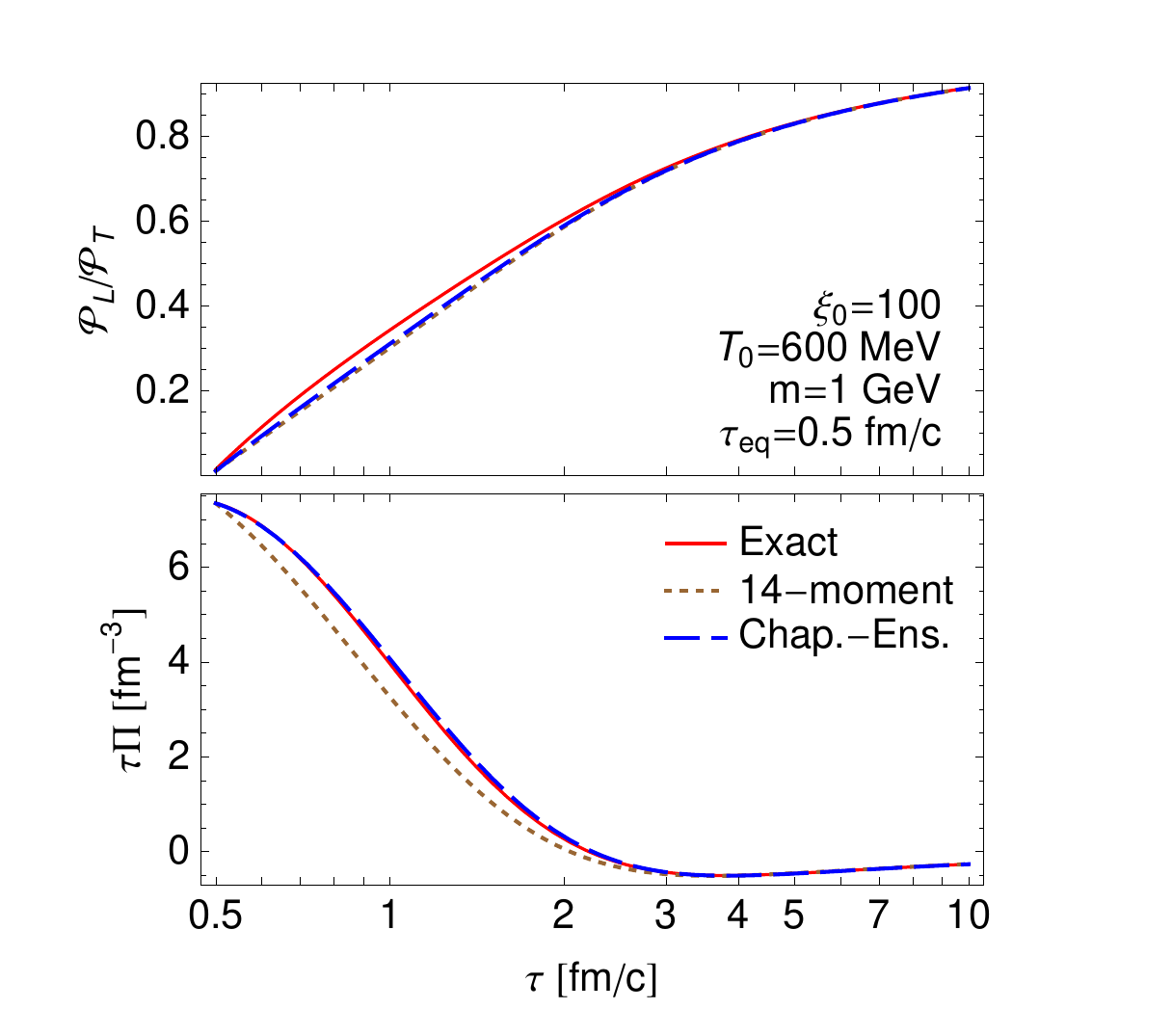}
\end{center}
\vspace{-0.8cm}
\caption{(Color online) Same as Fig. \ref{fig_100_03} except here 
	we take $m=1$ GeV which for $\xi_0 = 100$ implies 
	$\pi_0=35.12$ GeV/fm$^3$ and $\Pi_0=3.08$ GeV/fm$^3$.} 
\label{fig_100_1}
\end{figure}


\section{Boost-invariant (0+1)-dimensional case}

In the case of transversely homogeneous and purely-longitudinal 
boost-invariant expansion \cite{Bjorken:1982qr}, all scalar 
functions of space and time depend only on the longitudinal proper 
time $\tau=\sqrt{t^2-z^2}$. In terms of Milne coordinates, 
$(\tau,x,y,\eta)$, the hydrodynamic four-velocity becomes 
$u^\mu=(1,0,0,0)$. The energy-momentum conservation equation 
together with equations (\ref {BULK}) and (\ref{SHEAR}) reduce to
\begin{align}
\dot\epsilon &= -\frac{1}{\tau}\left(\epsilon + P + \Pi -\pi\right) \, ,  \label{epsBj}\\
\dot\Pi + \frac{\Pi}{\tau_\Pi} &= -\frac{\beta_\Pi}{\tau} - \delta_{\Pi\Pi}\frac{\Pi}{\tau}
+\lambda_{\Pi\pi}\frac{\pi}{\tau} \, ,  \label{bulkBj}\\
\dot\pi + \frac{\pi}{\tau_\pi} &= \frac{4}{3}\frac{\beta_\pi}{\tau} - \left( \frac{1}{3}\tau_{\pi\pi}
+\delta_{\pi\pi}\right)\frac{\pi}{\tau} + \frac{2}{3}\lambda_{\pi\Pi}\frac{\Pi}{\tau} \, , \label{shearBj}
\end{align}
where $\pi\equiv-\tau^2\pi^{\eta\eta}$. We note that in this case 
the term involving the vorticity tensor, $2\pi_{\gamma}^{\langle\mu} 
\omega^{\nu\rangle\gamma}$, vanishes and hence has no effect on the 
dynamics of the fluid. We also note that the first terms on the 
right-hand side of Eqs.~(\ref{bulkBj}) and (\ref{shearBj}), are the 
first-order terms $\beta_\Pi\theta$ and $2\beta_\pi \sigma^{\mu\nu}$, 
respectively, whereas the rest are of second-order.

We solve Eqs.~(\ref{epsBj})-(\ref{shearBj}) simultaneously assuming 
an initial temperature of $T_0=600$ MeV at the initial proper time 
$\tau_0=0.5$ fm/$c$, with relaxation times $\tau_{\rm eq}=\tau_\Pi= 
\tau_\pi=0.5$ fm/$c$ corresponding to $(\eta/s)_{\tau=\tau_0}=3/4\pi$. 
We solve the equations
for two different initial pressure configurations, $\xi_0=0$, 
corresponding to an isotropic pressure configuration $\pi_0=\Pi_0=0$ 
and $\xi_0=100$ corresponding to a highly oblate anisotropic 
configuration. Here $\xi$ is the anisotropy parameter which is 
related to the average transverse and longitudinal momentum in the 
local rest frame via $\xi=\frac{1}{2} \langle p_T^2\rangle/\langle 
p_L^2\rangle-1$. We consider two different masses, $m=300$ MeV 
roughly corresponding to the constituent quark mass and $m=1$ GeV 
representing the approximate thermal mass of a gluon or quark. For 
comparison, we also solve Eqs. (\ref{epsBj})-(\ref{shearBj}) with 
transport coefficients obtained by using the 14-moment method \cite 
{Denicol:2014vaa, Denicol:2014mca}.


In Figs. \ref{fig_0_03} -- \ref{fig_100_1} we show the proper-time 
evolution of the pressure anisotropy ${\cal P}_L/{\cal P}_T\equiv 
(P+\Pi-\pi)/(P+\Pi+\pi/2)$ (top) and the bulk viscous pressure times 
proper time (bottom) for three different calculations: the exact 
solution of the RTA Boltzmann equation \cite {Florkowski:2014sfa} 
(red solid line), second-order viscous hydrodynamics using the 
14-moment method \cite {Denicol:2014vaa} (brown dotted line), and 
the Chapman-Enskog method used herein (blue dashed line). Figures 
\ref {fig_0_03} and \ref{fig_100_03} show the case that $m=300$ MeV, 
while Figs. \ref{fig_0_1} and \ref {fig_100_1} show the case that 
$m=1$ GeV. Figures \ref {fig_0_03} and \ref {fig_0_1} correspond to an 
isotropic initial condition ($\xi_0=0$), while Figs. \ref 
{fig_100_03} and \ref {fig_100_1} correspond to a highly oblate 
anisotropic initial condition ($\xi_0=100$). 

From Figs. \ref{fig_0_03} -- \ref{fig_100_1}, we see that ${\cal 
P}_L/{\cal P}_T$ is quite insensitive to whether one uses the 
14-moment or Chapman-Enskog transport coefficients obtained herein. 
However, the result for $\tau\Pi$ using the Chapman-Enskog method is 
in better agreement with the exact solution of the RTA Boltzmann 
equation than the 14-moment method. 

\begin{figure}[t]
\begin{center}
\includegraphics[width=\linewidth]{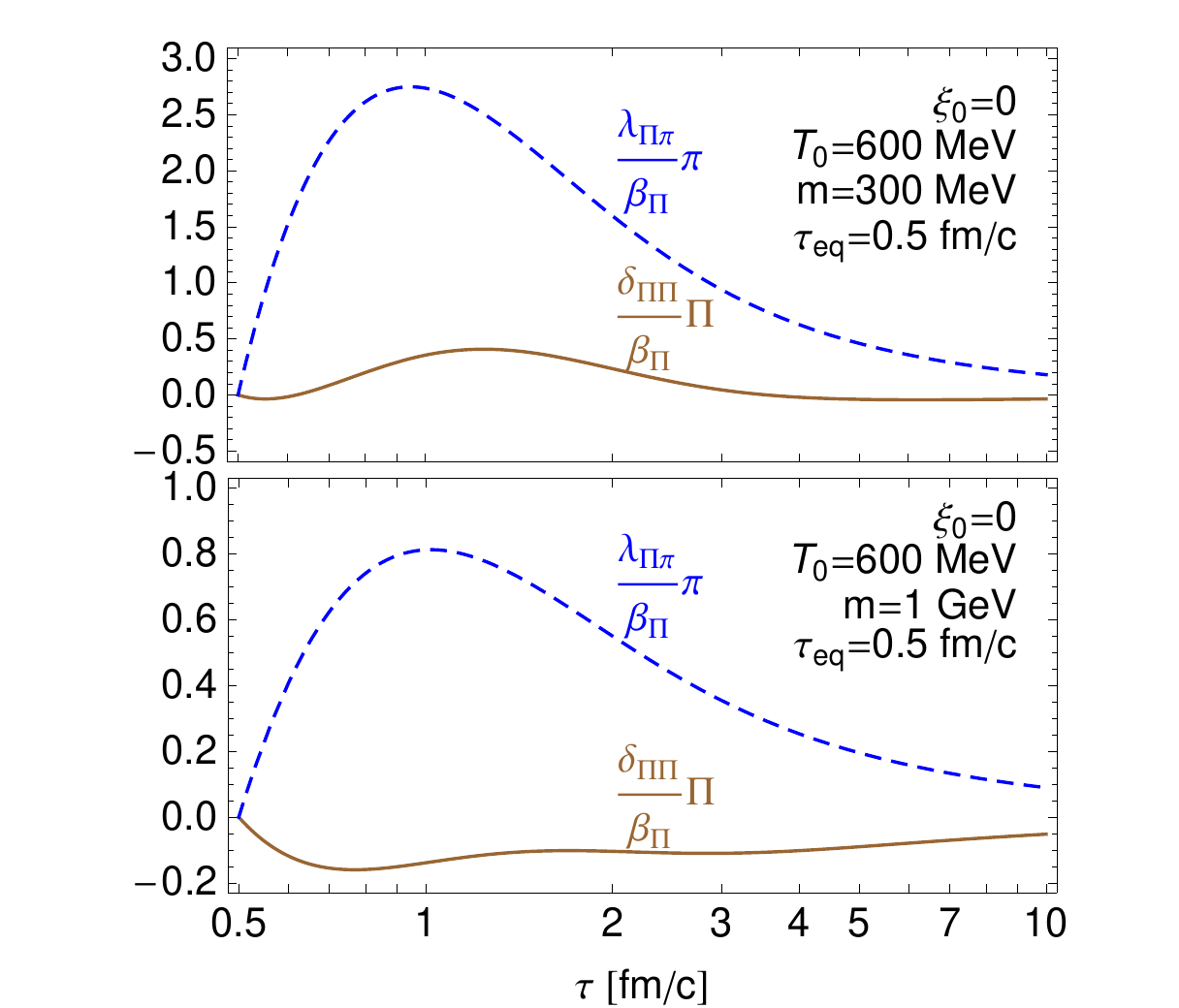}
\end{center}
\vspace{-0.45cm}
\caption{(Color online) Proper time evolution of the second-order terms 
	scaled by the first-order term in the evolution equation for 
	bulk viscous pressure, Eq. (\ref{bulkBj}). For both panels we 
	use $T_0=600$ MeV at $\tau_0=0.5$ fm/$c$, and $\tau_{\rm 
	eq}=\tau_\pi=\tau_\Pi=0.5$ fm/$c$. The initial spheroidal 
	anisotropy in the distribution function, $\xi_0=0$, corresponds 
	to an isotropic pressure configuration $\pi_0=0$ and $\Pi_0=0$. For 
	the top panel, we show results for $m=300$ MeV whereas the bottom 
	panel corresponds to $m=1$ GeV.} 
\label{fig_rel}
\end{figure}

In Fig. \ref{fig_rel} we plot the proper-time evolution of the 
second-order terms scaled by the first-order term in the evolution 
equation for bulk viscous pressure, Eq. (\ref{bulkBj}). We observe 
that for $m=300$ MeV (top panel), the relative magnitude of the 
shear-bulk coupling term is greater than unity for the proper-time 
interval $0.6\lesssim\tau\lesssim 3$ fm/$c$ indicating that the 
evolution of bulk viscous pressure is dominated by its coupling to the 
shear for a long time on the time scales relevant to hydrodynamic 
evolution in relativistic heavy-ion collisions. For the case of 
$m=1$ GeV (bottom panel), although the effect is not as 
prominent, the shear-bulk coupling term is still almost as 
important as the first-order expansion scalar. 


\section{Conclusions and outlook}

In this paper we applied the iterative Chapman-Enskog method to the 
derive second-order viscous hydrodynamical equations and the 
associated transport coefficients for a massive gas in the relaxation-time approximation. The resulting dynamical equations (\ref{BULK}) 
and (\ref{SHEAR}) have precisely the same form as those obtained 
using the 14-moment approximation \cite{Denicol:2014vaa}; however, 
some of the transport coefficients are different than those obtained 
in the 14-moment approximation when $m>0$. The equivalence or 
in-equivalence of the various transport coefficients was established 
analytically by using Taylor expansions in $m/T$ and also by direct 
numerical evaluation of the necessary integrals.  

Having obtained the full set of dynamical equations necessary to 
self-consistently evolve both the bulk pressure and shear tensor, we 
then specialized to the case of a transversally homogeneous and 
longitudinally boost-invariant system. In this specific case it is 
possible to solve the RTA Boltzmann equation exactly \cite 
{Florkowski:2014sfa}. Using this solution as a benchmark, we 
computed the pressure anisotropy and bulk pressure evolution using 
both the Chapman-Enskog method presented herein and the 14-moment 
method used in Ref.~\cite{Denicol:2014vaa}. We demonstrated that the 
Chapman-Enskog method is able to reproduce the exact solution better 
than the 14-moment method. For the pressure anisotropy both methods 
give very similar results, but for the bulk pressure evolution the 
Chapman-Enskog method better reproduces the exact solution. 

Finally, we presented a comparison of the magnitude of the 
shear-bulk coupling term in the dynamical equations for the bulk 
pressure to the term proportional to the first-order expansion 
scalar. We showed that, on the time scales relevant for relativistic 
heavy ion collisions, the shear-bulk coupling in the bulk pressure 
evolution equation is equally as important as the term involving the 
expansion scalar, in agreement with previous findings \cite 
{Denicol:2014mca}. We therefore conclude that once the second order 
terms for the bulk pressure are taken into account, at least in the 
relaxation time approximation, we obtain very good agreement with the 
exact solution of the RTA Boltzmann equation. Since the latter does 
not rely on order-by-order expansion of the distribution function 
about equilibrium, this can be taken as evidence that, in the RTA, the 
second-order terms capture the most important non-equilibrium 
corrections.

At this point, we would like to clarify that we are using the exact 
solution of the RTA Boltzmann equation as a benchmark to compare 
different hydrodynamic formulations and that our minimal requirement 
for a viable non-conformal hydrodynamic theory is that it should be 
able to describe the dynamics in this simple case.  It is true 
that the dynamics becomes more complicated when 
realistic scattering kernels are considered.  These could, in fact, lead to a completely 
different parametric behavior for bulk viscosity \cite{Jeon:1995zm,Lu:2011df}.  
Looking forward, since the shear-bulk 
coupling term is as important as the first-order term, we believe it would be 
interesting to determine its impact in higher dimensional 
simulations. Moreover, from a phenomenological perspective, a large 
negative bulk viscous correction might lead to early onset of cavitation. It 
would therefore be instructive to see how the second-order transport 
coefficients obtained here influence cavitation. In addition, it 
would also be interesting to see whether the second-order results derived 
herein could be extended to third order. We leave these questions 
for a future work. 


\acknowledgments{ 
We thank G. Denicol for useful discussions. R.R. was supported by 
Polish National Science Center Grant No.~DEC-2012/07/D/ST2/02125 and 
U.S.~DOE Grant No.~DE-SC0004104.   M.S. was supported in part by 
U.S.~DOE Grant No.~DE-SC0004104.
}


\end{document}